# Magnetic quantum phase transition in Cr-doped $Bi_2(Se_xTe_{1-x})_3$ driven by the Stark effect

Zuocheng Zhang[1,*], Xiao Feng[1,*], Jing Wang[2,3,*], Biao Lian[3], Jinsong Zhang[1], Cuizu Chang[1], Minghua Guo[1], Yunbo Ou[1], Yang Feng[1], Shou-Cheng Zhang[3,4], Ke He[1,4,†], Xucun Ma[1,4], Qi-Kun Xue[1,4], Yayu Wang[1,4,†]

[1]*State Key Laboratory of Low Dimensional Quantum Physics, Department of Physics, Tsinghua University, Beijing 100084, P. R. China*

[2]*State Key Laboratory of Surface Physics, Department of Physics, Fudan University, Shanghai 200433, P. R. China*

[3]*Department of Physics, Stanford University, Stanford, CA 94305–4045, USA*

[4]*Collaborative Innovation Center of Quantum Matter, Beijing, China*

*\* These authors contributed equally to this work.*

† Emails: kehe@tsinghua.edu.cn; yayuwang@tsinghua.edu.cn

The interplay between magnetism and topology, as exemplified in the magnetic skyrmion systems[1], has emerged as a rich playground for finding novel quantum phenomena and applications in future information technology. Magnetic topological insulators (TI) have attracted much recent attention, especially after the experimental realization of quantum anomalous Hall effect[2]. Future applications of magnetic TI hinge on the accurate manipulation of magnetism and topology by external perturbations, preferably with a gate electric field. In this work, we investigate the magneto transport properties of Cr doped $Bi_2(Se_xTe_{1-x})_3$ TI across the topological quantum critical point (QCP). We find that the external gate voltage has negligible effect on the magnetic order for samples far away from the topological QCP. But for the sample near the QCP, we observe a ferromagnetic (FM) to paramagnetic (PM) phase transition driven by the gate electric field. Theoretical calculations show that a perpendicular electric field causes a shift of electronic energy levels due to the Stark effect, which induces a topological quantum phase transition and consequently a magnetic phase transition. The *in situ* electrical control of the topological and magnetic properties of TI shed important new lights on future topological electronic or spintronic device applications.

The nontrivial bulk band topology in TI originates from strong spin-orbit coupling (SOC), which causes an inversion between the bulk conduction and valence bands[3-5]. When the $Bi_2Te_3$ family TIs are doped with transition metal elements such as Cr and V, long-range FM order can be established[6-9]. The broken time-reversal symmetry opens an energy gap at the Dirac point of the topological surface state, and the quantum anomalous Hall effect is realized when the Fermi level ($E_F$) lies within the bulk and surface band gaps[2,10-13]. The utilization of magnetic TI for device applications requires the *in situ* manipulation of its electronic and magnetic properties via a gate voltage, as has been achieved in conventional diluted magnetic semiconductors[14-17]. For magnetic TI, the gate electric field has two distinct effects on its physical properties. First, it injects charge carriers into the sample and can tune the position of $E_F$ across the Dirac point of the surface states. A series of experiments have been performed along this direction, and have revealed novel carrier-mediated magnetic order and transport phenomena that are unique to TIs[8,18,19]. Second, a perpendicular electric field within the sample can shift the electronic energy levels through the Stark effect, which may induce a topological phase transition[20-23] followed by magnetic phase transition[24]. However, the Stark effect induced topological-magnetic phase transition in TI has not been realized experimentally. The main reason is that the energy shift due to electric field is usually a small perturbation compared to the other characteristic energy scales in TI, such as the large SOC strength and bulk energy gap. Moreover, in order to reveal the consequence of the Stark effect, the carrier-dependent magnetic phenomena have to be excluded in the gate-tuning experiment.

These obstacles can be circumvented by an appropriate choice of magnetic TI materials. Here we grow 8 QL (quintuple layer) Cr doped $Bi_2(Se_xTe_{1-x})_3$ thin films on $SrTiO_3$ (STO) substrate by molecular beam epitaxy (MBE). The schematic TI-based field effect transistor device is shown in Fig. 1a, and details about the experiment can be found in the Methods session and supplementary materials. Figure 1b displays the anomalous Hall effect (AHE) curves measured at $T = 1.5$ K on 8 QL $(Bi_{0.89}Cr_{0.11})_2(Se_xTe_{1-x})_3$ with $x = 0$, 0.52, 0.67, 0.86 and 1 (temperature dependent AHE curves are shown in supplementary Fig. S1). For $x = 0$ and 0.52, the AHE curves show square-shaped hysteresis characteristic of FM order. For $x \geq 0.67$, however, the AHE curves are perfectly reversible, indicating the disappearance of long range FM order and transition into a PM state as discussed in a previous report[25]. The corresponding angle-resolved photoemission spectroscopy (ARPES) band maps shown in supplementary Fig. S2 reveal a topological phase transition by varying the Se/Te ratio. The Se substitution of Te reduces the SOC strength that was already weakened by Cr substitution of Bi. At sufficiently high Se content the SOC is too weak to invert the bulk bands, driving the system to a topologically trivial phase. This topological quantum phase transition in turn drives a magnetic phase transition via the van Vleck mechanism, which states that the FM order is favored only when the bulk bands are inverted[10,25]. The bulk band structure of the $x = 0.67$ sample is thus on the verge of inversion, as confirmed by the ARPES band map shown in Fig. 1c.

The 8 QL $(Bi_{0.89}Cr_{0.11})_2(Se_xTe_{1-x})_3$ thin films are thus perfect for realizing the Stark effect induced topological-magnetic phase transition. First, this system can be tuned close to a topological QCP so the Stark effect may be sufficient to drive a topological phase transition. Second, this system is heavily electron doped with carrier density in the order of $10^{14}/cm^2$ as estimated from the slope of the normal Hall effect in Fig. 1b, which is also consistent with the fact that $E_F$ lies deep in the bulk conduction band (Fig. 1c and supplementary Fig. S2). The carrier density induced by the applied gate voltage on the STO substrate is only in the order of $10^{13}/cm^2$ (Ref. 26 and supplementary section D), hence cannot cause significant change of the carrier dependent magnetism (see supplementary section E for details). More interestingly, the dielectric constant of STO ($\varepsilon_{STO}$ ~ 20000 at 1.5 K) is three orders of magnitude larger than that of the TI film ($\varepsilon_{TI}$ ~ 33 at 1.5 K). Therefore a small gate voltage applied on the STO substrate can produce a large out-of-plane electric field across the TI film[24].

Now we are ready to investigate the effect of the gate electric field on the magnetic order. We start from 8 QL $(Bi_{0.89}Cr_{0.11})_2Te_3$ without Se doping ($x = 0$). Shown in Fig. 2a are the AHE curves measured at $T = 1.5$ K under different $V_g$ between ±210 V. There is a well-defined hysteresis loop in all the AHE curves, and the coercive field ($H_C$) is almost a constant over the whole $V_g$ range. The gate electric field apparently has negligible effect on the FM order in this sample, except for a small change of the anomalous Hall resistance at weak field. Similar behavior can be found in the $x = 0.52$ sample (Fig. 2b), which is still in the Te rich topological and FM phase. When we increase the Se content to $x = 0.86$ (Fig. 2c), which is in the Se rich topologically trivial phase, the AHE curves are reversible without any observable hysteresis. There is only a negative weak-field curvature characteristic of a PM material[25]. With varied gate voltage, the PM behavior remains qualitatively the same and the AHE curves only show a small shift. The 8 QL $(Bi_{0.89}Cr_{0.11})_2Se_3$ sample without Te ($x = 1.0$) exhibits very similar behavior (Fig. 2d), a PM phase that is insensitive to the applied gate voltage.

In sharp contrast to these four samples, the magnetic properties of the sample near the topological QCP can be significantly changed by the gate voltage. As shown in Fig. 3a, the AHE curves measured at $T = 1.5$ K on the $x = 0.67$ sample display squared-shaped hysteretic loop at $V_g = -210$ V with a positive jump at zero field, which is characteristic of long-range FM order. With reduced gate voltage, the hysteresis disappears at $V_g = -25$ V. At the same time, the positive jump in AHE curves is also gradually reversed to a negative curvature. Supplementary Fig. S5 shows the corresponding Arrott plots, which can accurately determine the magnetic order and magnetic phase transition[27,28]. It is clear that the intercept at $V_g = -210$ V is positive while the intercept at $V_g = -25$ V is negative, which unambiguously demonstrates that a FM to PM phase transition is realized by changing the gate voltage. Further increasing the gate voltage to $V_g = +210$ V, the AHE curves remain almost the same. Similar magnetic quantum phase transition is observed at $T = 3$ K, as shown in Fig. 3b, although in general the FM order is weaker and exists in a smaller $V_g$ range.

When the temperature is increased to $T = 5$ K (Fig. 3c), the FM order disappears over the whole $V_g$ range.

The total two-dimensional (2D) Hall resistance in a FM material can be expressed as $R_{yx} = R_A M + R_N H$, where the first term is the AHE caused by spontaneous magnetization and the second term is the ordinary Hall resistance[29]. The spontaneous Hall resistance $R_{yx}^S$ can be extracted from the intercept of the Arrott plot of Hall resistance, which characterizes the spontaneous magnetization $M_S$ (Ref. 17,27,28). In the FM (PM) state, the intercept of the Arrott plot is positive (negative). The $R_{yx}^S$ is taken to be zero when the Arrott intercept becomes negative, indicating that the sample is in the PM state. The estimation of $R_{yx}^S$ at varied gate voltages before and after the subtraction of ordinary Hall resistance are shown in supplementary Session F. Figure 3d summarizes $R_{yx}^S$ in the $x = 0.67$ sample as a function of $V_g$ measured at $T = 1.5$ K. Both $R_{yx}^S$ before and after the subtraction of ordinary Hall resistance show positive values for $V_g \leq -100$ V and become zero for $V_g \geq -50$ V. The disappearance of both hysteresis loop and spontaneous Hall resistance $R_{yx}^S$ clearly demonstrate the magnetic phase transition tuned by external gate voltage. It should be noted that $V_g = -50$ V lies in the vicinity of the magnetic phase transition, and the zero field Hall resistance $R_{yx}^0$ still has a small positive value when the magnetic field is reduced from positive to zero. However, the Arrott intercept at $V_g = -50$ V (Fig. S6) is already negative and thus we take the spontaneous Hall resistance $R_{yx}^S$ as zero. At $V_g = -210$ V, $R_{yx}^S$ has a finite value at 3 K and becomes zero at 5 K, indicating that the Curie temperature ($T_C$) is between 3 K and 5 K (supplementary Fig. S6). Figure 3e displays the colored contour plot of the anomalous Hall resistance measured at $T = 1.5$ K, which separates the $H$-$V_g$ phase diagram into two distinct regimes. The positive AHE effect is always associated with the FM state while the negative AHE effect only exists in the PM regime.

Next we discuss the mechanism of the electric field driven magnetic phase transition observed here. Previous experiments on magnetic TI have also revealed FM to PM phase transition controlled by gate voltage, but the main effect there is the injected charges that change the position of $E_F$. It has been demonstrated that both surface and bulk charge carriers can mediate ferromagnetism via the Ruderman-Kittel-Kasuya-Yosida (RKKY) mechanism. In the current work, however, we can safely exclude the carrier dependent ferromagnetism (see supplementary section E for details). First, the 8 QL $(Bi_{0.89}Cr_{0.11})_2(Se_xTe_{1-x})_3$ films studied here are heavily electron doped with carrier density in the order of $10^{14}/cm^2$, so the surface Dirac fermions mediated ferromagnetism is negligible due to the large Fermi wavelength ($k_F \sim 1.22$ Å$^{-1}$)(Ref. 18,30). Second, the long range FM order shown in Fig. 3a is induced by negative gate voltage, which injects holes hence reduces the total electron-type carrier density. Therefore, the gate-voltage dependence of the FM order is totally opposite to that induced by bulk RKKY effect[8,19,31]. Third, the charges injected by gate voltage only cause a small variation of the total carrier density. This point is clearly illustrated by the ordinary Hall effect whose negative slope is insensitive to $V_g$ (supplementary Fig. S3). Moreover, the amount of injected charges is nearly the same for all samples, but the magnetic phase transition is not observed for Se content away from the $x = 0.67$ topological QCP. If the carrier dependent ferromagnetism could lead to the magnetic phase transition, all the samples would show similar electrically driven magnetic phase transition. These observations suggest that the injected charges themselves are not essential for the magnetic phase transition.

Another important fact is the striking similarity between the gate-tuned AHE curves in Fig. 3a and the Se doping dependent AHE curves in Fig. 1b. There is a FM to PM phase transition accompanied by the sign reversal of the AHE, while the ordinary Hall effect remains mostly intact. This similarity suggests that the gate electric field plays a similar role as the isovalent Se substitution of Te, i.e., the modification of effective SOC strength that affects the band topology, which in turn drives a magnetic phase transition.

These experimental observations lead us to propose that the Stark effect, the shift of energy levels due to external electric field, is the underlying mechanism for the electrically tuned magnetic phase transition reported here. The low energy physics of the TI system is described by the bonding state of $p_z$ orbit of Bi and the antibonding states of $p_z$ orbit of Se/Te, which are originally inverted due to the large SOC coupling[32]. A perpendicular electric field across the sample causes an opposite shift of the two orbits through the Stark effect, as schematically shown in Fig. 4a, which effectively reduces the band inversion. Under sufficiently strong electric field, these two orbits will become non-inverted, leading to a topological phase transition to a trivial phase, which tends to be PM due to the weakening of the van Vleck mechanism[10,25].

The Stark effect induced magnetic phase transition is further supported by the TI electronic structure obtained by theoretical calculations, as described in details in the Method session and supplementary materials. Although the top and bottom surfaces are effectively decoupled in the 8 QL TI film, for the sample at the verge of topological phase transition the top and bottom surface states will hybridize and open a surface gap at the Dirac point due to the reduced bulk band gap[33]. Effectively, these gapped surface states are no longer reside at the surfaces, but merge into the bulk. Therefore, in our theoretical calculations we consider 8 QL $x = 0.67$ sample as a quasi-2D system with 2D subbands (see supplementary section G for details). Figure 4b displays four representative band structures for a TI near the topological QCP subjected to different perpendicular electric fields. In the absence of electric field ($E = 0$), the bulk band structure is inverted with a sizable energy gap $\Delta \sim 19$ meV. A perpendicular electric field $E = 0.043$ V/nm causes a splitting and shift of the bands, and the inverted energy gap is reduced to $\Delta \sim 11$ meV. At

$E = 0.096$ V/nm, the bulk conduction and valence bands touch each other, and the sample now lies right at the topological QCP. Further increase of $E$ to 0.122 V/nm drives the sample to the topologically trivial regime and reopens a small normal band gap with $\Delta \sim 3$ meV. We use the capacitor model[24] to estimate the corresponding gate voltage $V_g = \frac{\varepsilon_{TI} d_{STO}}{\varepsilon_{STO}} (E - E_0)$, where $E$ is electric field and $E_0$ is the effective intrinsic band-bending field. In our calculations $E_0$ is taken to be zero and thus 0.122 V/nm electric field corresponds to $V_g = 50$ V on STO substrate. Figure 4c summarizes the calculated band gap as a function of the perpendicular electric field, which clearly illustrates the Stark effect induced topological phase transition. Figure 4d displays the calculated spin susceptibility $\chi_{zz}$ along the $z$-axis, which decreases rapidly with increasing electric field. The underlying physical picture is that the van Vleck spin susceptibility is much more pronounced in the topologically nontrivial regime than the topologically trivial regime, which will lead to the FM to PM phase transition.

Although the Stark effect can give a natural explanation for the observed phenomena, we notice that there is a shift between the calculated phase diagram (Fig. 4c,d) and the experimental one (Fig. 3d). A highly plausible cause is the existence of a perpendicular electric field $E_0$ within the sample even without an applied gate voltage. For the samples studied here, both band bending and charge transfer between the substrate and sample could build up this initial perpendicular electric field. The band bending effect has been revealed in previous experiments at the surface or interface in various TI systems[26,34-37], which can induce an electrical potential in the order of 60 mV between the top and bottom surfaces in 6 QL $Bi_2Se_3$ thin film[38]. Another particularly important issue is the oxygen vacancies in the STO substrate, which can transfer electron type carriers into the TI film[39,40]. The accumulation of $n$-type carriers at the bottom interface could also build up a perpendicular electric field within the TI film at $V_g = 0$. Under such circumstances a negative $V_g$ injects holes that neutralize the *n*-type charges at the interface and reduce the perpendicular electric field. According to the Stark effect picture, the topologically nontrivial inverted band structure is favored at negative $V_g$, which resolves the quantitative discrepancy between the theoretical calculation and experimental results. It should also be noted that the topological phase transition and magnetic phase transition may not happen at exactly the same external gate voltage.

The tuning of band gap by the Stark effect has also been realized in black phosphorous[41], transition metal dichalcogenides[42] and nanotubes[43,44]. The unique feature of the magnetic TI studied here is that the band gap is closely entangled with topology and magnetism, making it possible to tune the topological and magnetic properties simultaneously. Another advantage of the current system is the AHE can be used as a direct and convenient signature for the bulk band topology, which is otherwise difficult to probe without *in situ* ARPES measurements. The electrical control of topological order and magnetic properties of TI also facilitates the fabrication of electronic or spintronic devices based on the nature of topology[24,45], which are more robust against external perturbations such as disorder.

## Methods:

**Sample growth and transport measurements:** The $(Bi_{0.89}Cr_{0.11})_2(Se_xTe_{1-x})_3$ thin films with thickness of 8 QL are grown on 0.25 mm thick $SrTiO_3(111)$ substrate by using MBE and is fabricated into a field effect transistor device with the STO serving as the bottom gate dielectric. The film is covered with 20 nm Te capping layer before being taken out of the MBE chamber to avoid contaminations. The sample is scratched into a Hall-bar geometry and electrical contacts are made by mechanically pressed Indium. The magneto transport properties are measured by using standard four-probe ac lock-in method with the magnetic field applied perpendicular to the film plane.

**Theoretical calculations:** We first calculate the self-consistent electric field energy potential distribution by employing a Poisson-Schrodinger equation in the TI/STO system under various gate voltages (see supplementary for details). We then calculate the band structure of an 8 QL TI film under the z-axis electric field by using the tight-binding models. We take the effective 4-band $k \cdot p$ Hamiltonian $H_0$ of a $Bi_2(Se,Te)_3$ TI thin film near the topological QCP[24], and set the thickness of film to 8 QLs ~ 8 $nm$. The self-consistent iteration procedure is as follows: given an electric potential energy $V(z)$ where $0\ nm \leq z \leq 8\ nm$ is the out-of-plane direction coordinate inside the thin film, one can diagonalize the total Hamiltonian $H_0 + V$ to find all the electronic wave functions. Under a fixed $E_F$, the charge density distribution $\rho(z)$ can then be calculated by adding up probabilities of occupied electron wave functions at $z$. Solving the Poisson equation $\nabla^2 V = -\rho/\epsilon_{TI}\epsilon_0$ with this $\rho(z)$ in turn gives the potential energy $V(z)$. When the iteration procedure converges, we use the wave functions to calculate the spin susceptibility $\chi_{zz}$ with the perturbation theory for a TI with finite chemical potential.

## Acknowledgements

We thank Peizhe Tang, Jia Li and Haijun Zhang for helpful discussions. This work was supported by the National Natural Science Foundation of China and the Ministry of Science and Technology of China. B. L., J. W. and S.C. Z. acknowledge the support from the U.S. Department of Energy, Office of Basic Energy Sciences, Division of Materials Sciences and Engineering, under contract

No. DE-AC02-76SF00515. J. W. also acknowledges the support from the National Thousand-Young-Talents Program.

Additional information: The authors declare no competing financial interests.

Figure Captions:

**Figure 1 | Experimental setup and materials characterization. a,** Schematic drawing of a 8 QL $(Bi_{0.89}Cr_{0.11})_2(Se_xTe_{1-x})_3$ film grown on the $SrTiO_3$ substrate and fabricated into a field effect transistor device with a bottom gate. 20 nm amorphous Te is used for a capping layer. **b,** The AHE of 8QL $(Bi_{0.89}Cr_{0.11})_2Te_3$ shows square-shaped hysteresis at $T = 1.5$ K, which becomes smaller with increasing Se content and finally disappears at $x = 0.67$. The AHE curves become reversible with further increasing Se content to $x = 1$. **c,** ARPES band map of the $x = 0.67$ sample shows that the bulk conduction and valence bands almost touch each other, indicating the close proximity to a topological QCP. The Fermi level lies deep in the bulk conduction band due to the existence of electron type bulk carriers.

**Figure 2 | Gate-tuned AHE for samples far from the topological QCP. a,b,** In the well-defined topological nontrivial and FM phase, gate electric field has negligible effect on the FM order in $x = 0$ (a) and $x = 0.52$ (b) samples except for the small change in anomalous Hall resistance at low magnetic field. **c,d,** The $x = 0.86$ (c) and $x = 1$ (d) samples in the topologically trivial regime remain PM regardless of the applied gate voltages. The AHE curve only shows a small shift with applied gate voltage.

**Figure 3 | Gate-tuned magnetic phase transition near the topological QCP. a-c,** Temperature dependent AHE curves at varied gate voltages for the $x = 0.67$ sample. The hysteresis disappears at $V_g = -25$ V for $T = 1.5$ K (a) and at $V_g = -50$ V for $T = 3$ K (b). The anomalous Hall curves are reversible for the whole $V_g$ range for $T = 5$ K (c). **d,** The spontaneous Hall resistance before and after the subtraction of ordinary Hall resistance clearly reveal the gate-tuned magnetic phase transition at $T = 1.5$ K. **e,** Contour plot of the total anomalous Hall resistance as a function of $V_g$ and magnetic field at $T = 1.5$ K. Positive AHE is associated with FM order while negative AHE exists in the PM regime.

**Figure 4 | The Stark effect induced topological phase transition. a,** Schematic plot of the energy level shift caused by the Stark effect, in which a perpendicular electric field reduces the inverted band gap. **b**, Calculated bulk band structures at four representative electric fields. The inverted band gap at $E = 0$ decreases with increasing electric field, and closes at $E = 0.096$ V/nm. Further

increase of electric field drives the band structure to topological trivial regime with a normal band gap. **c,** The calculated band gap as a function of electric field. The arrow indicates the critic electric field where the topological phase transition occurs. **d,** The calculated magnetic susceptibility decreases rapidly with increasing electric field and almost saturates when the sample lies in the topologically trivial regime.

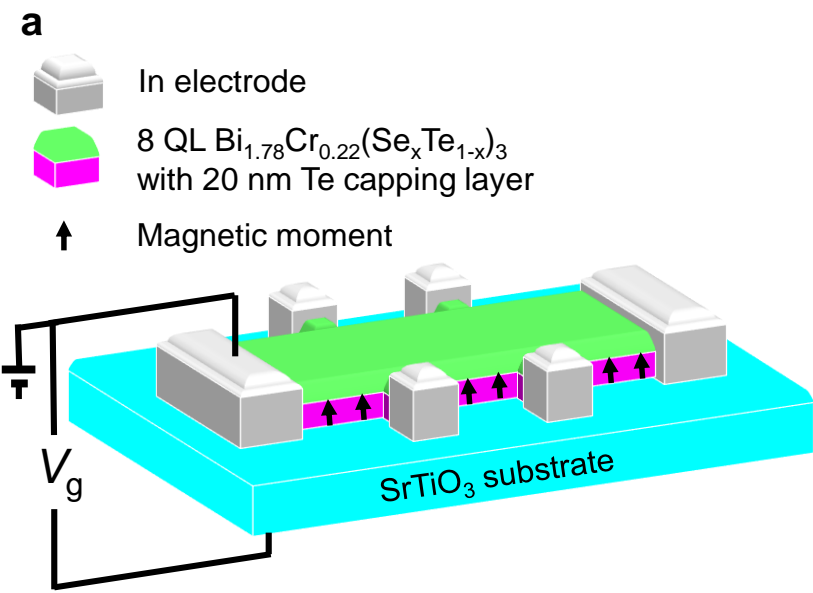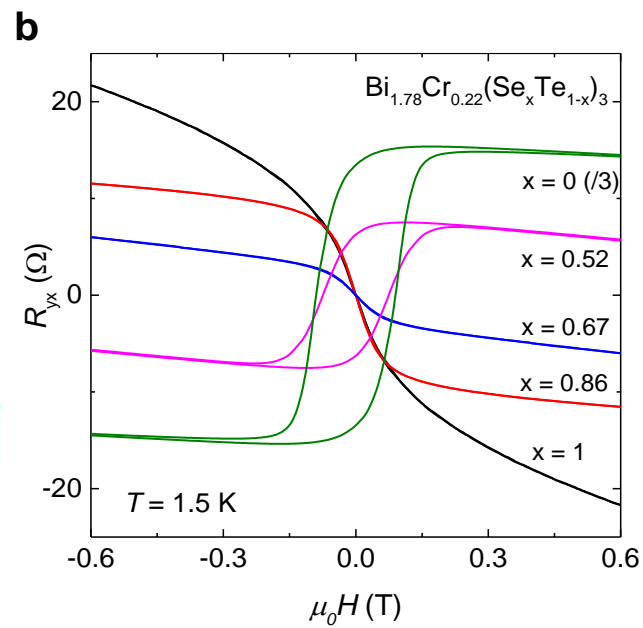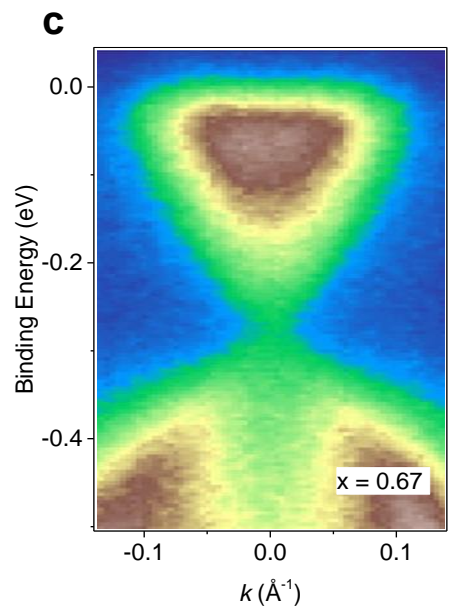

**Figure 1**

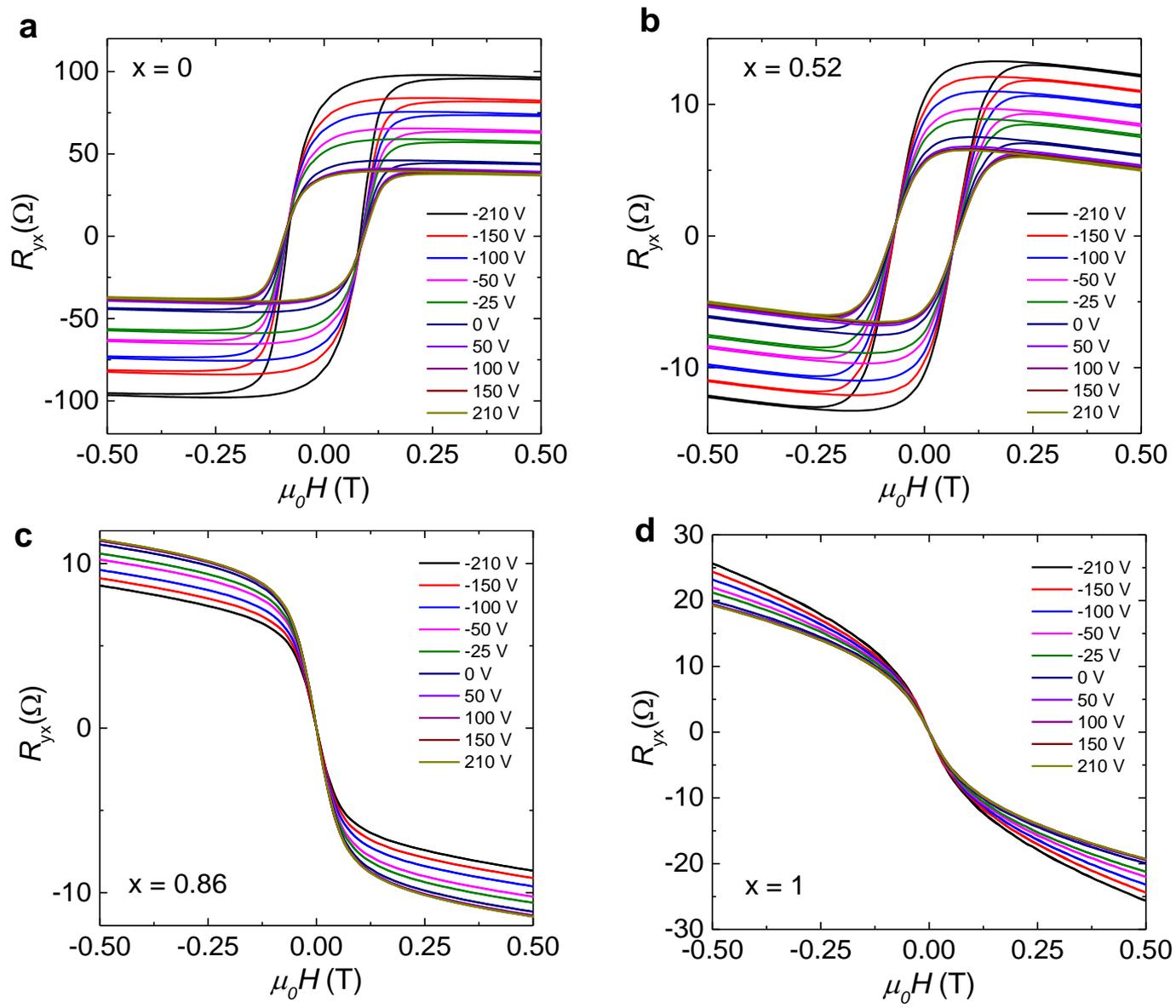

**Figure 2**

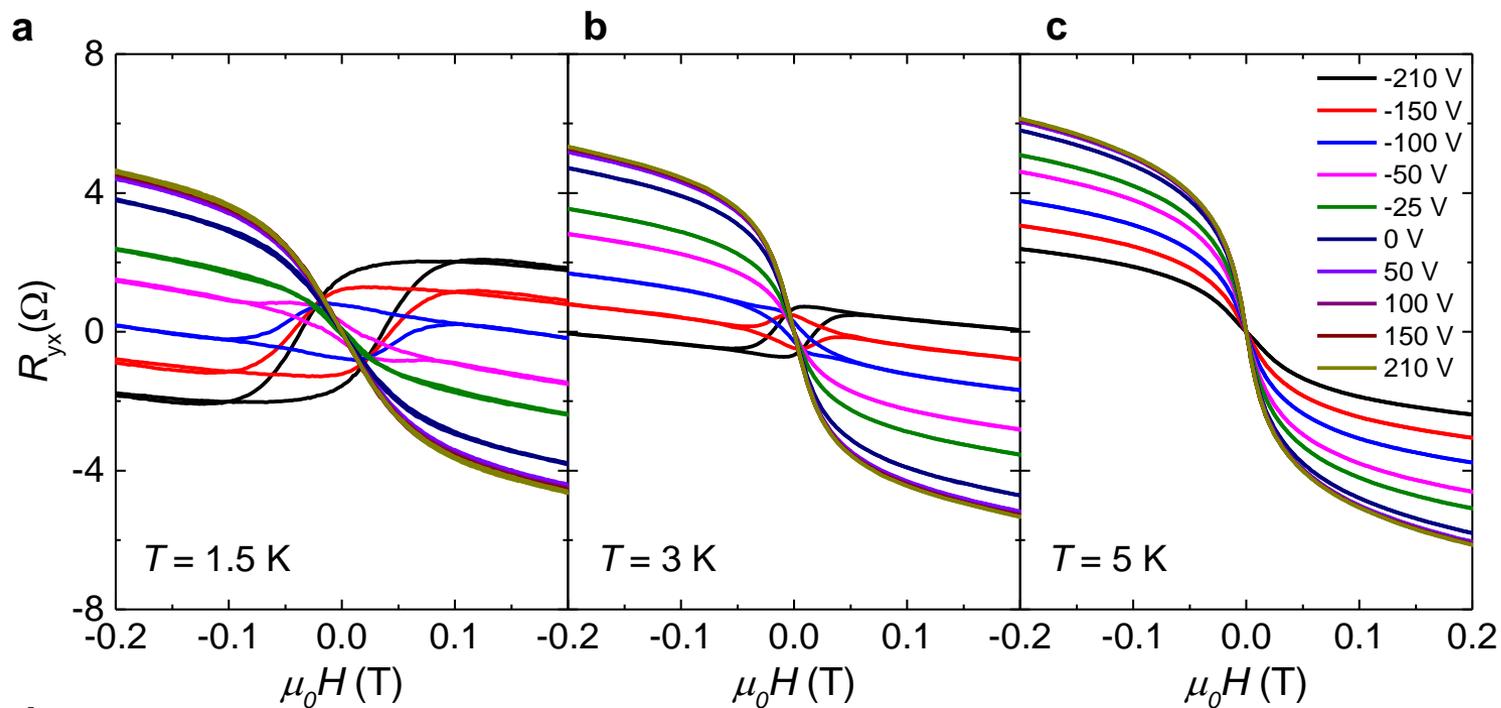

**Figure 3**

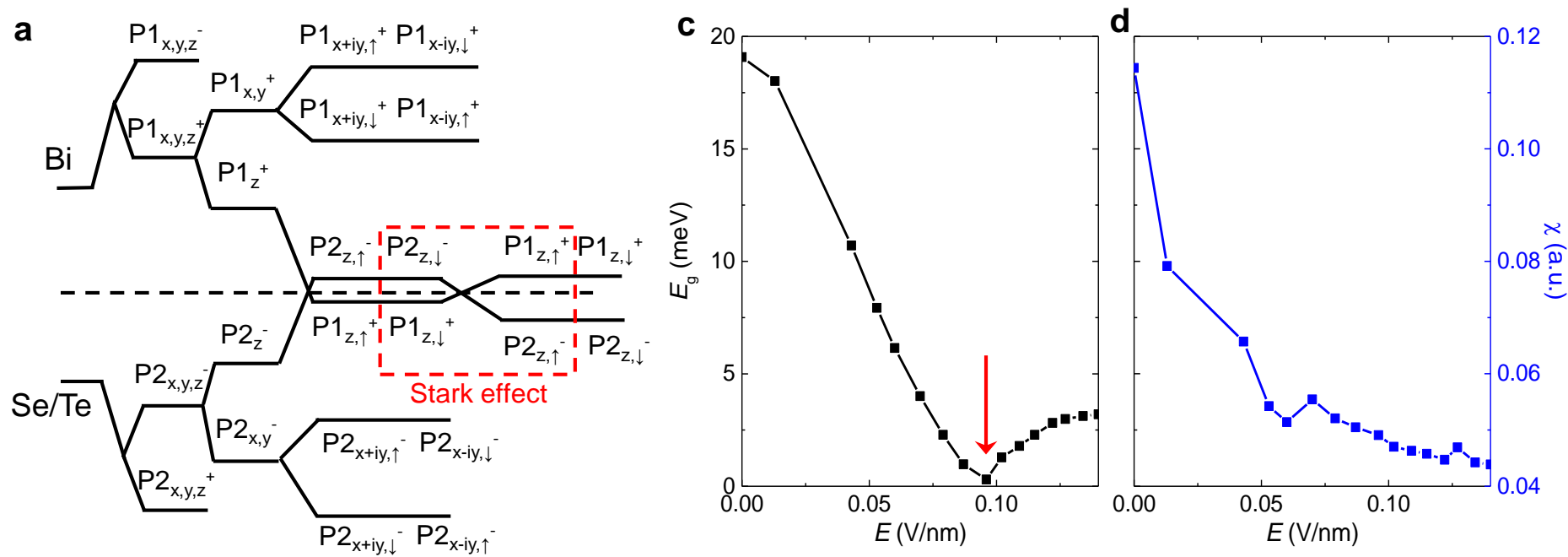
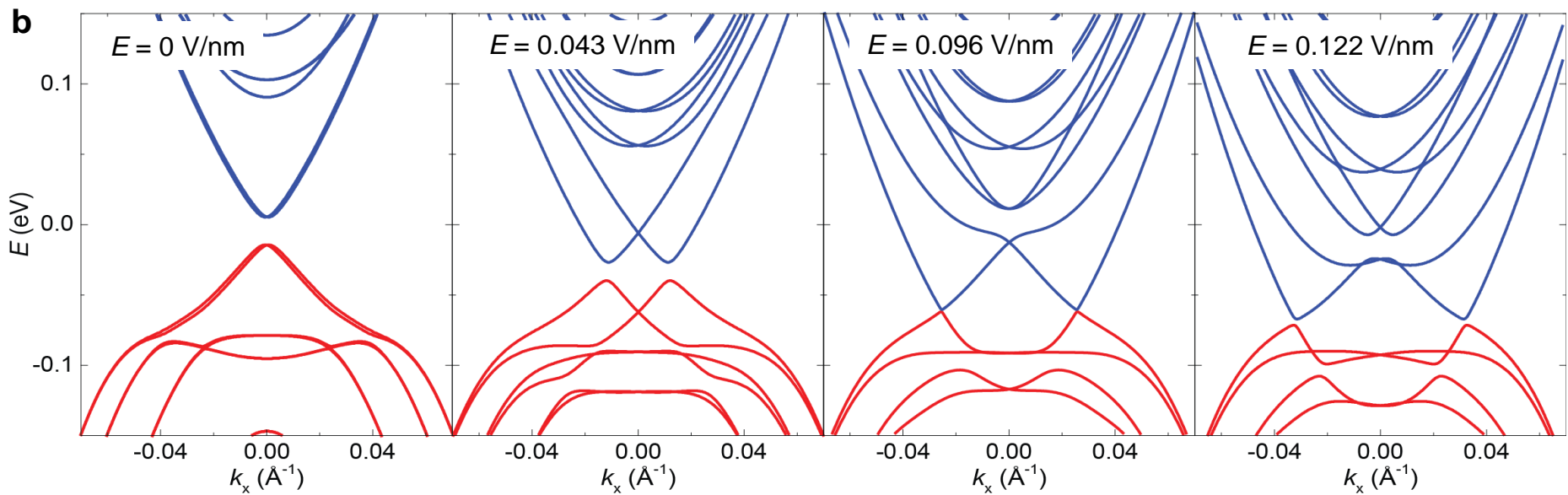

**Figure 4**